\documentclass[draft]{agujournal_mod}

\begin{document}

%
%

\title{Amplitude of solar wind density turbulence from 10--45 $R_{\odot}$}

%
%

 \authors{K. Sasikumar Raja\affil{1}\thanks{Indian Institute of Science Education and Research, Pashan, Pune - 411 008},
 Madhusudan Ingale\affil{1}\thanks{Currently at Physical Research Laboratory, Navrangpura, Ahmedabad - 380 009}, R. Ramesh\affil{2},
 Prasad Subramanian\affil{1}\thanks{Also Centre for Excellence in Space Sciences, India. http://www.cessi.in}, P. K. Manoharan\affil{3} and P. Janardhan\affil{4}}

\affiliation{1}{Indian Institute of Science Education and Research, Pashan, Pune - 411 008}
\affiliation{2}{Indian Institute of Astrophysics, 2nd Block, Koramangala, Bangalore - 560 034}
\affiliation{3}{Radio Astronomy Centre, National Centre for Radio Astrophysics, Tata Institute of Fundamental Research, Udhagamandalam (Ooty) - 643 001}
\affiliation{4}{Physical Research Laboratory, Navrangpura, Ahmedabad - 380 009}


\correspondingauthor{K. Sasikumar Raja}{sasikumar@iiserpune.ac.in}


\begin{keypoints}
\item How do the solar wind density modulation index and turbulence amplitude vary with heliocentric distance and sunspot number?
\item Is the large scale coronal magnetic field linked to solar wind density turbulence?
\end{keypoints}

%
%

\begin{abstract}
We report on the amplitude of the density turbulence spectrum ($C_{N}^{2}$) and the density modulation index ($\delta N/N$) in the solar
wind between $10$ and $45 R_{\odot}$. 
We derive these quantities using a structure function that is observationally 
constrained by occultation observations of the Crab nebula made in 2011 and 2013 and similar observations published earlier. We use the most general form of the structure function, together with currently used 
prescriptions for the inner/dissipation scale of the turbulence spectrum.  Our work yields a comprehensive 
picture of a) the manner in which $C_{N}^{2}$ and $\delta N/N$ vary with heliocentric distance in the solar wind and b) of the solar cycle dependence of these quantities.

\end{abstract}

\section{Introduction}\label{intro}
The extended solar corona and the solar wind is a rich testbed for studying the properties of magnetohydrodynamic (MHD) turbulence. While most 
solar wind turbulence theories only treat incompressible turbulence, density irregularities are present in the solar wind, and are manifested through 
fluctuations in the refractive index. Knowledge of density turbulence impacts our understanding of the solar wind significantly, 
and is important for interpreting a variety of observations. It is linked to our basic understanding of the 
compressibility of solar wind turbulence (e.g., \citet{Tu1994,Hnat2005}).
It is also key to infer radio wave scattering leading to depressed quiet Sun brightness temperatures at low radio
frequencies \citep{The92,Sas1994,Ram00a,Sub2004,Ram06,The08}, the dissipation of solar wind turbulence, leading to extended solar wind heating (e.g., \citet{Car2009,Cha2009}), 
energetic particle propagation through the heliosphere (e.g., \citet{Rei2010}) and other interesting problems.

Density turbulence in solar wind has been studied using techniques ranging from angular broadening of radio 
sources (e.g., \citet{Arm90, Jan93,Ana94, Bas94, Spa95,Ram01a}) to spectral broadening \citep{Col89},
phase scintillations \citep{Woo1979}, interplanetary (intensity)
scintillations (IPS) \citep{Hew1964, Coh1969,Eke1971,Ric90,Man00,Bis2009,Bis2010,Tok2012,Tok2016} due to celestial radio sources and spacecraft radio beacons \citep{Ima2014}. 
Despite this impressive body of work, there are still
significant gaps in our understanding. For instance: while the spatial spectrum of density turbulence is 
generally acknowledged to follow the Kolmogorov scaling at relatively large scales, there is 
evidence for flattening of the spectrum near the inner/dissipation scale (e.g., \citet{Col89, Col91}). The 
location of the inner/dissipation scale is also a subject of considerable uncertainty. Another important 
quantity of interest is the so-called density modulation index $\delta N/N$, where $\delta N$ represents the turbulent density 
fluctuations and $N$ represents the background solar wind density. There have been some past attempts at measuring this quantity \citep{Woo95,Bav95,Spa02} and a relatively recent comprehensive study for 
heliocentric distances $> 40 R_{\odot}$ using the IPS technique \citep{Bis14b}.

Some of the uncertainties in our understanding of solar wind density turbulence are manifested in the 
debate regarding the smallest observable source in the solar corona at radio wavelengths. Since coronal turbulence broadens the 
source size, observations of compact sources place limits on the spectral amplitude of density turbulence. Observations reported by \citet{Lan87,Zlo92,Mer06,Mer15}  
at $\approx$ 327 MHz with angular resolutions $< 10$ arc sec
suggest that the smallest coronal radio source size is $\geq 30$ arc sec. 
Source sizes estimated from majority of the high angular 
resolution observations at lower frequencies ($\approx$ 30-100 MHz) also seem to be limited to $\geq 60$ arc sec
\citep{Wil98,Ram99,Ram00b,Ram01b,Ram12, Mug2016}, consistent with the predicted minimum observable
source sizes in this frequency range \citep{Rid74,Cai04}. However, much smaller coronal radio sources have also been 
reported at $\approx$ 170 MHz \citep{Ker79,Kat11}. Generally, the consensus is that scatter-broadened source sizes in the solar 
corona are most likely $\geq 10$ arc sec at 20 cm \citep{Bas94} and $\geq 3$ arc min at 100 MHz \citep{Bas04}.  This therefore emphasizes the need for reliable 
estimates of the amplitude of density turbulence at these scales, especially as a function of heliocentric distance. 

\begin{figure}[!ht]
\centering
\centerline{\includegraphics[width=12cm]{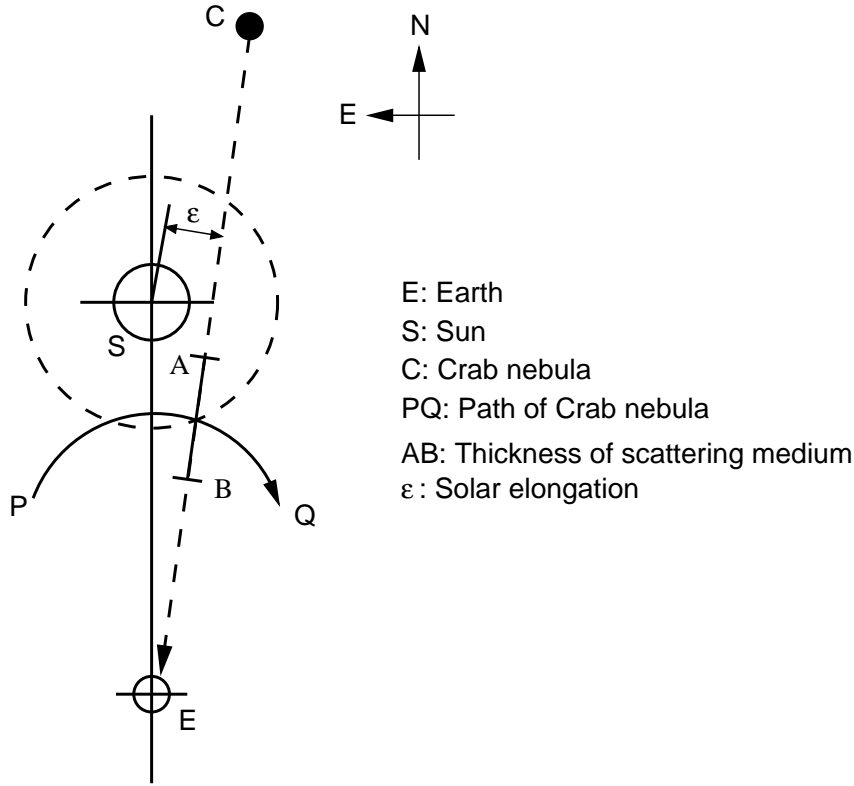}}
\caption{The schematic diagram shows the geometry of Crab nebula occultation; `PQ' indicates the 
projected path of the Crab nebula during the month of June. The closest point of `PQ' to `S' is $\approx$ $5~R_{\odot}$. 
The radiation from the `C' passes through the effective turbulent medium `AB' at a solar elongation of `$\epsilon$' as viewed from `E'.}
\label{fig:occult}
\end{figure}

In this work, we will use interferometric observations of the Crab nebula to infer the spectral level of solar wind density 
turbulence and the density modulation index as a function of heliocentric distance. Crab occultation is a very well established technique that has been in use since the 
1950s \citep{Hew57, Hew58, Hew63, Eri1964,Sas74}, giving us the advantage of a standard observational quantity to draw inferences 
from. The schematic diagram of the occultation is shown in Figure \ref{fig:occult}.
This technique is also best suited for turbulence density estimates in the $\approx 10-50 ~R_{\odot}$ heliocentric distance range.
The IPS technique at low frequencies usually probes heliocentric distances  $> 40-50 ~R_{\odot}$. IPS observations at microwave frequencies
probe the inner solar wind \citep{Eke1971, Yam1998, Ima2014}. Nonetheless, extensive studies of density turbulence amplitude and density modulation index and their solar cycle dependence were still lacking.

We have used Crab occultation observations made in 2011 and 2013 at the Gauribidanur observatory \citep{Ram11}, together with published data 
from several earlier observations by \citet{Mac52, Hew57, Hew58, Hew63} 
over the interferometer baselines $60-1000$ meters and frequencies $26-158$ MHz. We have scaled these measured structure 
functions to a baseline of 1600 meters and a frequency of 80 MHz, which were the parameters corresponding to Crab occultation observations in 2011 and 2013.

\section{Density turbulence: some background}

Turbulent density inhomogeneities in the solar corona 
are typically characterized by their spatial power spectrum
\begin{equation}
P_{\delta N}(k) = C_{N}^{2}(R) k^{-\alpha} e^{-(kl_{i}(R)/2\pi)^{2}}\, ,
\label{eq1}
\end{equation}
where $k$ is the (isotropic) wave number, and $l_{i}(R)$ is the
inner (dissipation) scale, where the spectrum steepens. The quantity $C_{N}^{2}$ 
is the amplitude of density fluctuations, and has dimensions of ${\rm cm}^{-\alpha - 3}$. 
There are not many estimates of $C_{N}^2$ in the literature; for 
example $C_N^2(R)$ is estimated using in-situ observations of Helios \citep{Mar90} and VLBI observations \citep{Sak93,Spa95,Spa96}. 
Using VLBI observations of phase scintillations, \citet{Spa95,Spa96} empirically quantified the dependence of $C_{N}^{2}$ on heliocentric distance as
\begin{equation}
C_{N}^{2}(R) = 1.8 \times 10^{10}(R/10)^{-3.66}
\label{eq2}
\end{equation}
over $R \approx 10-60~R_{\odot}$. They assumed a Kolmogorov spectrum ($\alpha = 11/3$)
for the density fluctuation, and the units of $C_{N}^{2}$ in Eq~\ref{eq2} are $m^{-20/3}$. We note that the spatial scales of the density inhomogeneities probed using VLBI are $\approx 200-2000$ km, 
which are substantially larger than the scales we are interested in ($\leq$ 10 km). To the best of our knowledge, our work provides the only parametrisation of the density turbulence amplitude as a function of heliocentric distance since \citet{Spa95,Spa96}.

Another important quantity of interest to us is the magnitude of the turbulent density fluctuations $\delta N_{k_i}$ at the inner scale ($l_i$), which can be related to the spatial power spectrum (Eq~\ref{eq1}) as follows \citep{Cha2009}:
\begin{equation}\label{eq:deltn}
{\delta}N_{k_i}^2(R) \sim 4 \pi k_i^3 P_{\delta N} (R, k_i) = 4 \pi C_{N}^{2}(R) k_i^{3 - \alpha} e^{-1} \, ,
\end{equation}
where we have used $k_{i} \equiv 2 \pi/l_{i}$.
Eq~(\ref{eq:deltn}) can be used to calculate the density modulation index $\epsilon_N(R)$ defined as
\begin{equation}\label{eq:df}
\epsilon_N(R) \equiv {~\delta N_{k_{i}}(R) \over N(R)} \, ,
\end{equation}
where $N$ is the solar wind background density. 

\section{Observations, structure function and the scattering measure}\label{sf}
We now briefly describe the Crab occultation observations and detail how we obtain the structure function and scattering measure from the measurements. 
These quantities will be used to compute $C_{N}^{2}$ (Eq~\ref{eq1}) and $\epsilon_{N}$ (Eq~\ref{eq:df})
\subsection{Crab occultation observations}

Since the Crab occultation technique is a well established one, we only briefly mention the aspects essential to our purpose. The Crab nebula 
is usually observed with a single element interferometer as it passes through the solar wind from 
$\approx$ 10--45 $R_{\odot}$ during mid June of every year. As it gets close to the Sun, its angular size increases due to enhanced scattering 
by the solar wind turbulent density irregularities. Eventually, its size increases to such an extent that it gets resolved out by the interferometer; 
the interferometer visibility decreases to unobservable levels, causing it to appear ``occulted''.

\begin{figure}[!ht]
\centering
\includegraphics[width=14cm]{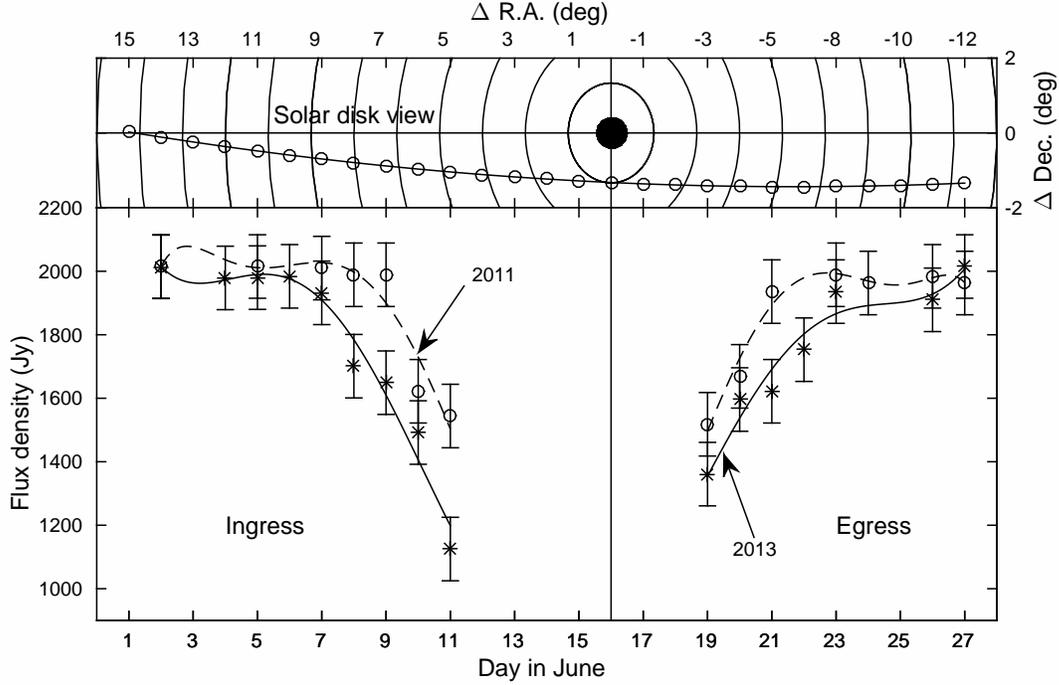}
\caption{The upper panel shows solar disk view of the Crab nebula occultation. 
The filled circle indicates the Sun and open circles represent the position of Crab nebula with respect to the Sun on different dates; 
$\Delta$ R.A. and $\Delta$ Dec are the offset distances of Crab nebula from the Sun in right ascension and declination respectively. 
The closest concentric circle around the Sun has a radius of $5~R_{\odot}$ and the radii of the rest of the circles differ from their 
adjacent ones by $5~R_{\odot}$. The bottom panel shows the observed flux densities of the Crab nebula on different days during its occultation by the
solar corona. The periods before and after June 16th correspond to the ingress and egress, respectively. The
plots marked `o' and `*' correspond to measurements during June 2011 and June 2013, respectively.}
\label{fig:figure2}
\end{figure}

The lower panel of Figure \ref{fig:figure2} shows the variation in the observed flux density of the Crab nebula during June 2011 and 2013;
while the upper panel shows the solar disk view of the occultation geometry.
While there is a steady decrease in the observed flux density (from the pre-occultation value
of $\approx 2015 \pm 100$ Jy) from 10th June 
($R \approx 23~R_{\odot}$) during the ingress 
in 2011, the decrease is noticeable from 8th June onwards ($R \approx 30~R_{\odot}$) 
in 2013 (see Figure \ref{fig:figure2}). 
A similar situation occurs during the egress.
While the pre-occultation value is reached
around 21st June ($R \approx 21~R_{\odot}$) in 2011, it is only around 23rd June 
($R \approx 29~R_{\odot}$) in 2013. No fringes were observed 
during $12-18$ June in both 2011 and 2013. 
The distance of the line-of-sight to the Crab nebula from the Sun,
was $R \approx 15~R_{\odot}$ on 12th June (ingress) and was $R \approx 10~R_{\odot}$ on 18th June (egress). 
Considering the fact that
the heliographic latitudes encountered by the Crab nebula during the ingress and egress 
are different \citep{Kun65}, we find that the occultation curves for 
the years 2011 and 2013 
in Figure \ref{fig:figure2} are fairly symmetric.
This is expected since 
the maximum of the solar cycle 24 was in the year 2013 and it has been shown that distribution 
of solar wind density fluctuations 
is spherically symmetric close to the solar maximum \citep{Man93}. 

In the present work, we have used these observations as well as similar ones made earlier. 
Crab nebula occultation observations were reported by \citet{Mac52} in 1952 at 38 and 80.5 MHz. Similar observations during 1952-1958 were reported in \citet{Hew57, Hew58}. 
These observations were made at 38, 81 and 158 MHz over baselines ranging from 60 - 1000 meters. 
Crab nebula occultation observations at 26.3 and 38 MHz over the baselines of $\approx$ 700 - 1630 meters were made during 1961 and 1962 by \citet{Hew63}. 
Furthermore, the 
normalized visibilities from the earlier observations, which were observed over different baselines and frequencies,
are used after scaling them to 80 MHz and a baseline of 1.6 km using the general structure function discussed in \S~\ref{sf1}.

The primary observational quantity inferred from the Crab nebula occultation technique is the visibility $V(s)$, which is essentially
the correlation between the voltages recorded by a pair of antennas. The visibility is a function of the observing baseline $s$. 
We will work with a quantity called the normalized visibility defined as $\Gamma(s) = V(s)/V(0)$. 
The structure function $D_{\phi}(s)$ which characterizes the phase perturbations caused by the density inhomogeneities in the medium is defined as \citep{Pro75,Ish78,Col89,Arm90}
\begin{equation}
\Gamma(s) = e^{-D_{\phi}(s)/2} \, .
\label{eq4}
\end{equation}
In other words,
\begin{equation}
D_{\phi}(s)=-2 ln \Gamma(s)=-2ln\left[V(s)/V(0)\right] \, ,
\label{eq3}
\end{equation}
where $V(s)$ and $V(0)$ are the ensemble averaged values. For our purposes, $V(0)$ 
is the flux density of the Crab nebula when it was far from the Sun. Crab occultation observations are typically made using a single baseline; i.e., one value of $s$. 

\subsection{The General Structure Function (GSF)}\label{sf1}

Over the years, theoretical developments and observations have converged on a well accepted formulation for the structure function to describe 
density fluctuations in the solar wind (e.g., \citet{Col87, Arm00, Bas94, Pra11}). 
These expressions for the structure function, however, are valid only for situations where the baseline $s$ is  $\ll$ the inner scale $l_i(R)$ 
or is $\gg$ the inner scale. These approximations might not hold in our situation; for 
(depending upon the inner scale model one assumes) there are situations where the observing baseline $s$ might be comparable to the inner scale. If this is the case, using the 
asymptotic expressions for the structure function will yield inaccurate results, and it is necessary to use the General Structure Function (GSF)
that is valid for the asymptotic regimes $s \ll l_{i}(R)$ and $s \gg l_{i}(R)$ and also straddles the intermediate regime $s \approx l_{i}(R)$ \citep{Ing14}.
Scatter-broadened images of sources observed against the background of the solar wind are observed to be anisotropic only for heliocentric 
distances $\leq 5$--6 $R_{\odot}$ \citep{Ana94, Arm90}. 
Since our observations are made for distances ranging from 10 to 45 $R_{\odot}$, it is adequate to use the isotropic GSF, which is defined as follows:
{\begin{eqnarray}
\label{gsf}
{D_\phi(s)} & ={{8 \pi^2 r_e^2 \lambda^2 \Delta L} \over {2^{\alpha-2}(\alpha-2)}} {\Gamma \bigg( 1 - {{\alpha-2} \over 2} \bigg)}
	    {{C_N^2 (R) l_i(R)^{\alpha-2}} \over {(1 - f_p^2 (R) / f^2)}} \\ \nonumber
	    & {\times \bigg\{ { _1F_1} {\bigg[ - {{\alpha-2} \over 2},~1,~ - \bigg( {s \over l_i(R)} \bigg)^2 \bigg]} -1 \bigg\}} \, ,
\end{eqnarray}}
where ${ _1F_1}$ is the confluent hypergeometric function, $r_e$ is the classical electron radius, 
$\lambda$ is the observing wavelength, $R$ is the heliocentric distance, $\Delta L$ is the thickness of the scattering medium,
$f_p$ and f are the plasma and observing frequencies respectively. The functional form of the structure function is thus well known;
the visibilities from the Crab occultation observations will provide one point that constrains its amplitude. The functional form of 
the structure function depends explicitly on the observing wavelength and the baseline. We use this dependence to normalise visibilities 
from Crab occultation observations made at different observing frequencies and wavelengths and baselines to an observing frequency of 80 MHz and a baseline of 1.6 km.

The origin of the inner (dissipation) scale $l_i(R)$ is a subject of intense ongoing research. While some researchers identify the inner scale
with the proton inertial length \citep{Col89,Har89,Yam98,Ver96,Lea99,Lea00,Smi01,Bru14}, some use the 
proton gyroradius for the inner scale \citep{Bale2005,Sah13, Bis14b}. These inner scale prescriptions are widely used in the literature, 
and we outline them in the \S \ref{AppA} for completeness. There are several instances where the
baseline lengths for the observations we consider are comparable to the inner scale. As shown in \S \ref{AppA}, the baseline length used in the 2011 and 2013
observations ($s = 1600$ meters) is comparable to the proton gyroradius for the relevant heliocentric distance range ($\approx 10$ -- $45 R_{\odot}$).
However, if the proton inertial length prescription is used for the inner scale, the typical baseline lengths are far 
smaller than the inner scale. We use the GSF (Eq~\ref{gsf}), which covers all these situations; it is accurate for $s \ll l_{i}(R)$ through $s \approx l_{i}(R)$ and extending to $s \gg l_{i}(R)$. 

\subsection{Inner scale models}\label{AppA}

{In this section we evaluate the inner scales in the solar wind using two different physical prescriptions and compare it with a fiducial interferometric baseline of 1600 meters.}
\subsubsection{Proton inertial length}

The mechanism of proton cyclotron damping by Alfv\'en waves is often invoked to account for the dissipation scale of solar wind turbulence. 
The inner scale predicted by this mechanism is the proton inertial length ($d_i$), \citep{Col89,Har89,Yam98,Ver96,Lea99,Lea00,Smi01,Che14,Bru14} which can be written as

\begin{equation}
 d_i(R) = 228 \times N_e(R)^{-1/2} \,\,\,\, {\rm km},
\label{protoninertial}
\end{equation}

\noindent where $N_e(R)$ is the background plasma density at heliocentric distance $R$ in ${\rm cm}^{-3}$. In order to calculate the 
background solar wind density, we start with daily peak values of the solar wind density at 1 AU during June 2011 and 2013 are used which were 
obtained from the Low Resolution OMNI (LRO) data set \footnote{http://omniweb.gsfc.nasa.gov/form/dx1.html}. 
For the rest of the years, the background solar wind density at different heliocentric distances $R$ (here, in units of AU) is extrapolated sunwards using the 
scaling predicted by the density model of \citet{Leb98}:

\begin{equation}
N(R) = 7.2~R^{-2} + 1.95 \times 10^{-3}~R^{-4} + 8.1\times 10^{-7}~R^{-6} \,\,\,\, {\rm cm}^{-3}.
\label{leblanc}
\end{equation}

Equation~\ref{leblanc} assumes a density of $7.2 \,\, {\rm cm} ^{-3}$ at 1 AU. To derive the background density at a specified $R$,
equation \ref{leblanc} is multiplied by N(1 AU)/7.2, where N(1 AU) denotes the peak value of the background 
density from the LRO data set. For the rest of the years the default Leblanc density model is used.
 
\begin{figure}[!ht]
\centering
\includegraphics[width=14cm]{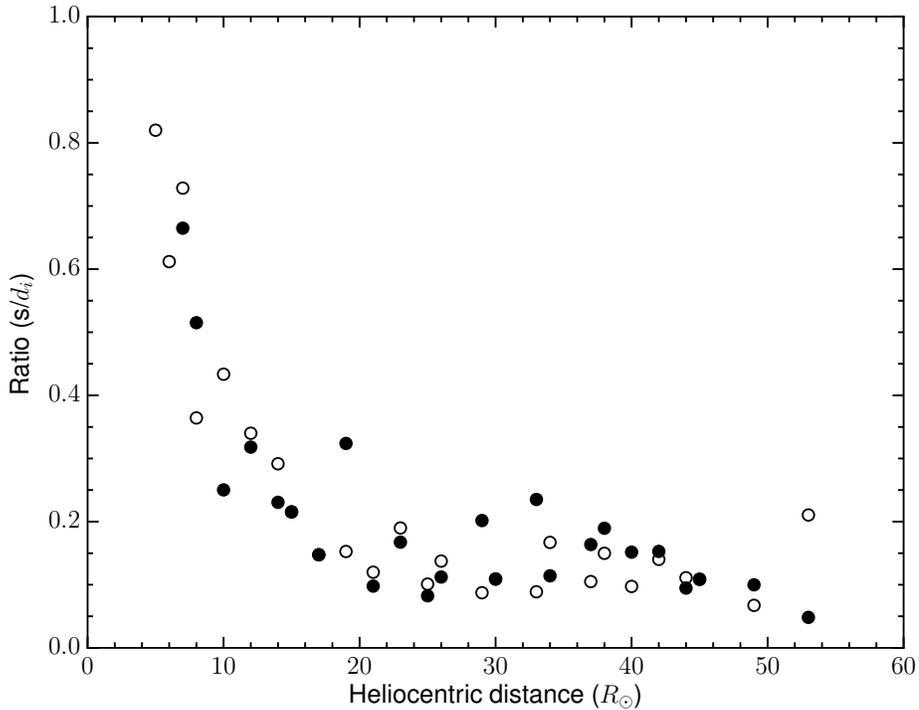}
\caption{Scatter plot of the ratio of the baseline to proton inertial length ($d_i$) plotted 
against heliocentric distance. The open and filled circles denotes the data points
derived using observations in June 2011 and 2013 respectively.
}
\label{inertial}
\end{figure}

Figure \ref{inertial} shows the ratio of the interferometric baseline used in June 2011 and 2013 (s = 1600 meters) to the proton inertial length. 
The open and filled circles correspond to the data points derived using observations in 
June 2011 and 2013 respectively. While $s$ is comparable to the inner scale for $R < 10 ~R_{\odot}$, it is significantly smaller than the 
inner scale for $R > 10~ R_{\odot}$. Since our data spans $45~ R_{\odot} > R > 10~ R_{\odot}$, it follows that the $s \ll l_{i}$ asymptotic 
branch is adequate if the inner scale is described by the proton inertial length.

\subsubsection{Proton gyroradius}
Another popular prescription for the inner/dissipation scale is the proton gyroradius \citep{Bale2005,Sah13, Bis14b, Che14}:
\begin{equation}
\rho_i(R) = 1.02 \times 10^2 \mu^{1/2} T_i^{1/2} B(R)^{-1} \,\,\,\, {\rm cm},
\label{ion_gyro}
\end{equation}

where $\mu (\equiv m_i/m_p)$ is the mass of an ion, in units of the proton mass, 
$T_i$ is the proton temperature in eV and $B$ is the interplanetary magnetic field in Gauss. 
In order to estimate the magnetic field, we begin with the daily average interplanetary 
magnetic field (IMF) at 1 AU from the LRO data set during June 2011 and June 2013. In order 
to obtain the IMF at a given heliocentric distance 
$R$, we extrapolate these values Sunward using the Parker spiral magnetic field model in the ecliptic plane \citep{Williams1995}:

\begin{equation}
B(R) = 3.4 \times 10^{-5}R^{-2}(1+R^2)^{1/2} \,\,\, {\rm Gauss},
\label{imf}
\end{equation}

\noindent where $R$ is the heliocentric distance in units of AU. This equation assumes a magnetic field of $= 4.7\times 10^{-5}$ Gauss at 1 AU.
We multiply equation \ref{imf} with B(1 AU)/($4.7\times 10^{-5}$), 
where B(1 AU) denotes the daily average IMF (in Gauss) obtained from the LRO data. Parker spiral magnetic field model is used 
as it is for the years other than 2011 and 2013. The inner scale lengths are calculated using equation \ref{ion_gyro} by assuming a proton temperatures of $T_i = 10^5$ K.

In the slow solar wind (300-400 km/s) the proton temperature would be $\approx 1 \times 10^5-6 \times 10^5$ K for heliocentric distances ranging 
from $\approx 0.2-0.05$ AU (i.e. $45-10 ~R_{\odot}$).
In the fast solar wind (700-800 km/s), the proton temperature would be $\approx 1.5 \times 10^6$ K at these heliocentric distances \citep{Mar91}.
The proton gyroradius for a proton temperature of $1.5 \times 10^{6}$K is $\approx 60~\%$ larger than that for a proton temperature of $10^{5}$K. 
We show the modulation index using a proton temperatures $10^5$ K as well as $1.5 \times 10^6$ K in 
Figures \ref{fig:dne_n} and \ref{fig:epsilon}.

\begin{figure}[!ht]
\centering
\includegraphics[width=14cm]{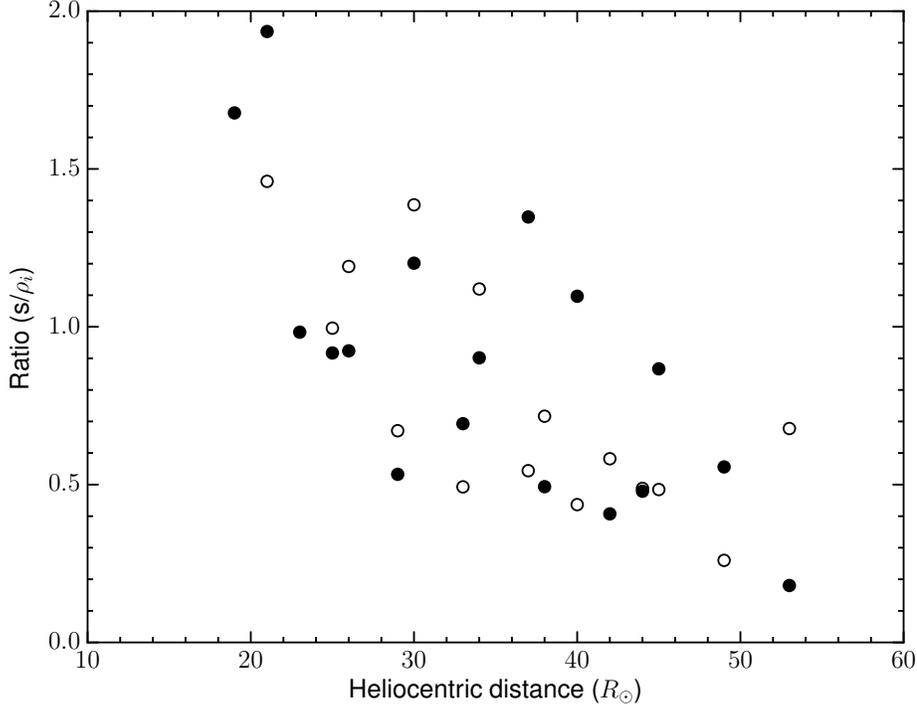}
\caption{Scatter plot of the ratio of the baseline to proton gyroradius ($\rho_i$ ) plotted
against heliocentric distance.  The `open' and `filled' circles denote the data points derived
using observations in June 2011 and 2013 respectively for the proton temperature $T_i = 10^5 $ K.}
\label{gyroradius}
\end{figure}

Figure \ref{gyroradius} shows the ratio of the interferometric baseline used for the June 2011 and 2013 observations 
(s = 1600 meters) to the proton gyroradius given by Eq \ref{ion_gyro}. The open and filled circles
denote the ratio corresponding to the observations in June 2011 and 2013 respectively for the proton 
temperatures $T_i = 10^5$ K. Evidently, $s \approx l_i(R)$ for $10$--$45~ R_{\odot}$, which is the
heliocentric distance range of interest to us. If the inner scale is the proton gyroradius, neither
of the asymptotic approximations $s \ll l_i(R)$ or $s \gg l_i(R)$ is therefore appropriate, and the GSF needs to be used.

\subsection{Estimating the scattering measure}

The scattering measure (SM) is defined as the path integral
\begin{equation}
\label{smdef}
{\rm SM} = \int C_{N}^{2}(R) \, dl \approx C_{N}^{2}(R) \, \Delta L\, ,
\end{equation}
where the integration is carried out over the depth over which scattering takes place. When the scattering is confined to a thin screen,
the approximation indicated in Eq~\ref{smdef} is acceptable, where $\Delta L$ is the thickness of the scattering screen.
We use the GSF defined in Eq~\ref{gsf} to calculate the scattering measure, which in turn will be used to determine $C_{N}^{2}(R)$. Accordingly,
\begin{eqnarray}\label{eq:sm1}
{\rm SM} &=& C_{N}^2(R) \Delta L \\ %
	 &=& \left(f(\alpha,\lambda) \, \frac{l_i(R)^{\alpha-2}}{r_f(R,\lambda)} \, \left(_1F_1(\alpha,s,R)-1\right) \right)^{-1} D_{\phi}(s) \nonumber
\end{eqnarray}

where,
\begin{eqnarray*}
f(\alpha, \lambda) &=& \frac{8\pi^2r_e^2\lambda^2}{2^{\alpha-2}(\alpha-2)} \Gamma\left(1-\frac{\alpha-2}{2}\right), \\
r_f(R,\lambda) &=& 1-f_p^2(R)/f^2, \\
_1F_1(\alpha,s,R) &=& _1F_1\left[-\frac{\alpha-2}{2},1,-\left(\frac{s}{l_i(R)}\right)^2 \right],
\end{eqnarray*}

\section{Results}\label{results}
\subsection{Heliocentric dependence of $C_{N}^{2}$}
As explained in \S \ref{sf}, the structure function $D_{\phi}(s)$ can be computed from the basic observed quantity $V(s)$. In turn, the structure function can be used 
to calculate the scattering measure (Eq~\ref{eq:sm1}). We now describe how the SM can be used to estimate the turbulence amplitude $C_N^2(R)$ at different 
solar elongations $R_{0}$ (which correspond to different observation dates in June, and therefore to different heliocentric distances).

Assuming solar wind turbulence at these heliocentric distances ($10-45~R_{\odot}$) to be spherically 
symmetric, the SM can also be expressed as \citep{Spa95}:

\begin{equation}
{\rm SM} = \int_0^\infty \,\, C_N^2(R)\,\, {\rm d}R = \frac{\pi}{2}C_{N}^2(R_{0})\, R_{0}
\end{equation}

\begin{equation}\label{eq:sm2}
C_{N}^2(R_{0}) = \frac{2}{\pi}\frac{\rm SM}{R_{0}}
\end{equation}
 where $C_{N}^2(R_{0})$ denotes the amplitude of density turbulence at impact parameter $R_0$. The impact parameter $R_0$ is related to 
 the solar elongation (see \S 52 in \citet{Duf11}) by a fraction $\approx 60/16$, where 16 is the solar radius in arc minute. Comparing with 
 Eq~\ref{smdef} shows that the scattering screen thickness is identified as
$\Delta L = (\pi/2) R_0$, in computing $C_{N}^2(R_{0})$ from the scattering measure.

The SM is estimated from the observed structure function ($D_{\phi}(s)$) using two inner scale models: the proton inertial length and and proton gyroradius 
(see \S \ref{AppA}). We use a proton temperature of $10^{5}$ K in the proton gyroradius prescription. Furthermore, 
the SM depends upon the assumed value of power 
law index ($\alpha$) of the density fluctuation spectrum (Eq~\ref{eq1}). Generally, the spectrum
is observed to follow a Kolmogorov-like scaling with $\alpha = 11/3$.
However, there is also some evidence for local flattening of the density fluctuation spectrum at large wave numbers \citep{Cel1987,Col89, Bas94};
some authors therefore use $\alpha = 3$. In view of the lack of consensus on this issue, we compute the 
SM for $\alpha = 11/3$ as well as $\alpha = 3$. Subsequently, $C_{N}^{2}$ is calculated from the SM using Eq~\ref{eq:sm2}.

\begin{figure}[!ht]
\centering
\includegraphics[width=12cm]{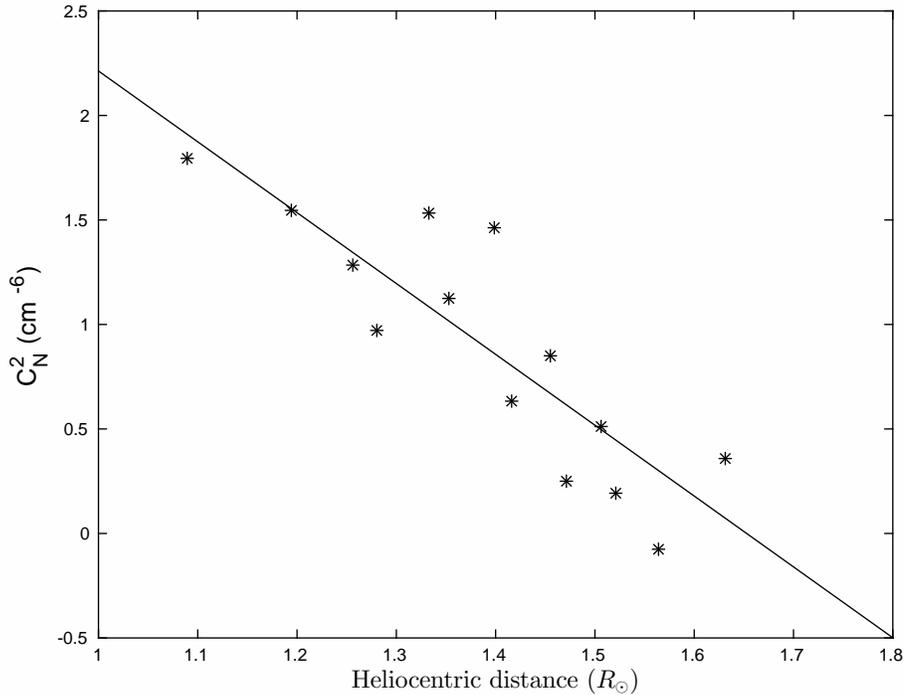}
\caption{A log-log scatterplot of $C_{N}^{2}$ against heliocentric distance (in $R_{\odot}$) derived from observations in 2013. We use $\alpha=3$ and
the inner scale is the proton inertial length. The fit to $C_N^2(R)=A~R^{-\gamma}$ yields $\gamma = 3.4$ and $A = 4 \times 10^{5}~{\rm cm}^{-6}$.}
\label{fig:cn21}
\end{figure}

Using all the available data described in \S~2, we compute $C_{N}^2$ as a function of heliocentric distance between 10 and 45 $R_{\odot}$. Since the observation span years corresponding
to solar minimum as well as solar maximum, we have studied the data from each year separately. For instance, Figure \ref{fig:cn21} shows 
the variation of $C_N^2$ with heliocentric distance 
using data from 2013.
We fit a function of the form $C_N^2(R)=A~R^{-\gamma}$ to the data plotted in these Figures. We find that the data in Figure \ref{fig:cn21} 
suggests $A = 4\times 10^5~{\rm cm}^{-6}$ and $\gamma = -3.4$ with a goodness of fit (adjusted $R^2$) 0.72.
Since we have a total of 44 such plots, we only show one representative example in Figure \ref{fig:cn21}, and tabulate all our results in 
Table \ref{tab:one}. It summarizes the heliocentric variation of $C_{N}^{2}$ for two values of $\alpha$ (11/3 and 3) and two inner scale models (the 
proton inertial length and the proton gyroradius). For instance, in 2011, $C_{N}^{2}(R) = 3.2 \times 10^{4} R^{-2.8}$ for $\alpha = 3$ and the proton 
inertial length as the inner scale. On the other hand, $C_{N}^{2}(R) = 400 R^{-2.1}$ for $\alpha = 3$ and the proton gyroradius (with proton temperature 
= $10^5$ K) as the inner scale. Table \ref{tab:one} is thus a comprehensive representation of the heliocentric distance dependence of $C_{N}^{2}$ between 
10 and 45 $R_{\odot}$. To the best of our knowledge, the only such result in the literature so far is due to \citet{Spa95} and \citet{Spa96}, who determined the 
heliocentric dependence of $C_{N}^{2}$ from 10 to 60 $R_{\odot}$ using VLBI observations during July and August 1991, which is $\approx$ 2 years past 
the maximum of cycle 22 in the declining phase. Their result, which assumes a Kolmogorov spectrum ($\alpha = 11/3$) is $C_{N}^{2}(R) = 3.81 R^{-3.66}$ 
in units of ${\rm cm}^{-20/3}$; the same result is quoted in a slightly different form in Eq~\ref{eq2}. Of the results we have compiled, data from 1960
corresponds to a similar phase in cycle 19. For this epoch, we obtain $C_{N}^{2} \propto R^{-\gamma}$, with $\gamma$ ranging from 3.2 to 3.3. Our results 
thus yield a remarkably similar dependence of $C_{N}^{2}$ with heliocentric distance for the only instance in the published literature where such a comparison can be made.

\subsection{Solar cycle dependence of $C_{N}^{2}(R)$}

\begin{figure*}
\centering
\includegraphics[width=12cm]{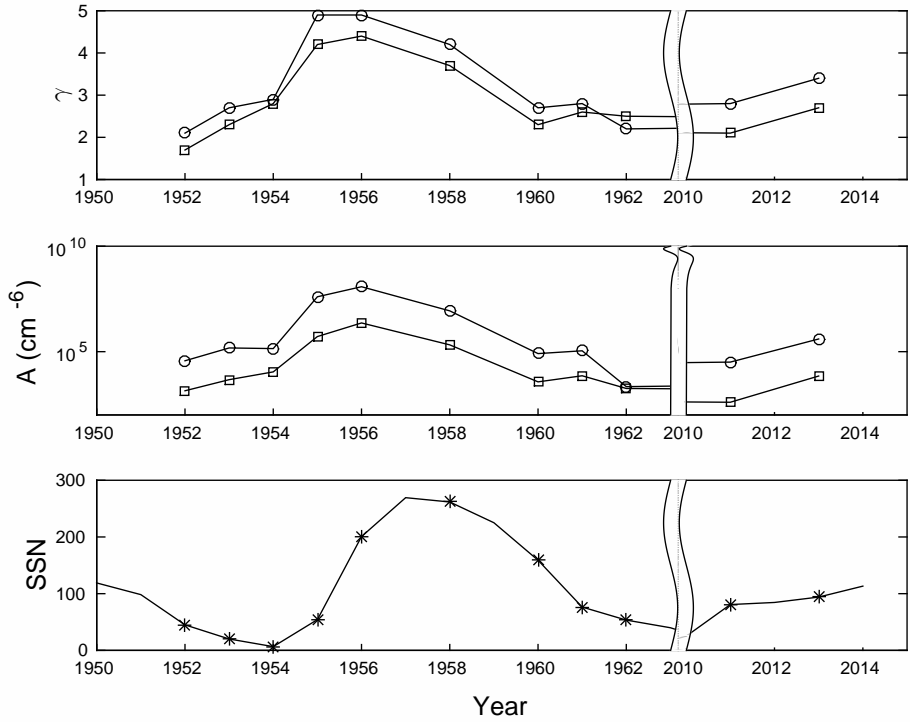}
\caption{The top panel and middle panel show $\gamma$ and $A~(cm^{-6})$ respectively as a function of time. 
The `circles' and `squares' represent the proton inertial and proton gyroradius inner scale models respectively with $\alpha=3$. For the proton gyroradius model, we use a temperature of $10^5$ K. 
The solid line in the bottom panel shows the yearly averaged sunspot number and the `*' represents 
the year in which the Crab occultation measurements were made. It is evident that both $\gamma$ and $A$ correlate well with the solar cycle.}
\label{fig:solarcycle_3}
\end{figure*}

\begin{figure*}\centering
\includegraphics[width=12cm]{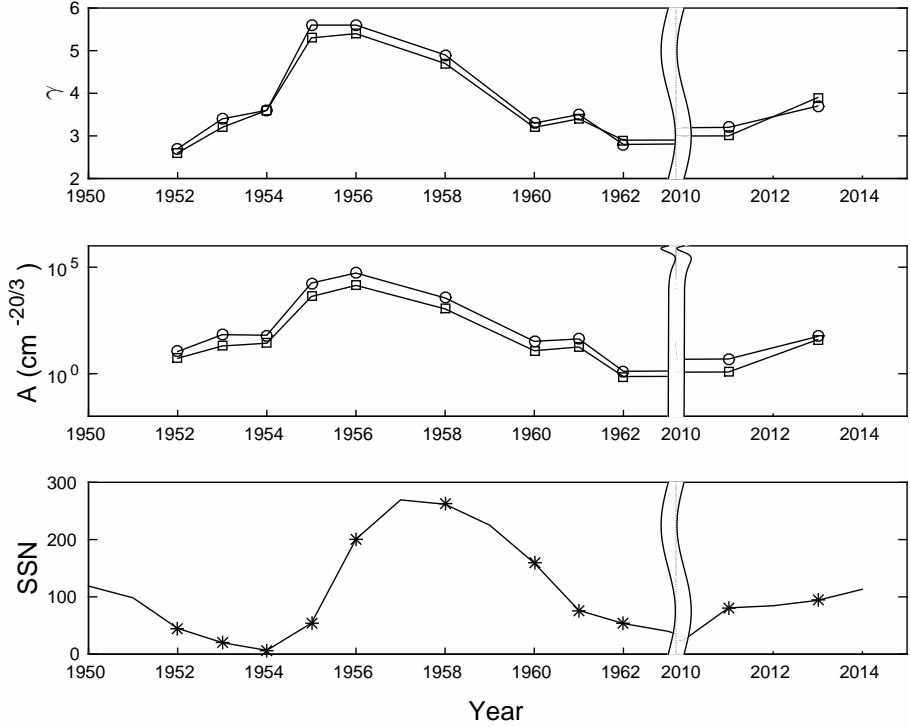}
\caption{Same as Figure~\ref{fig:solarcycle_3}, except for $\alpha$, which is 11/3. The dimensions of $A$ are therefore $cm^{-20/3}$.}
\label{fig:solarcycle_113}
\end{figure*}

It is evident from Table \ref{tab:one} that the values of $A$ and $\gamma$ are significantly different for different observation years, which correspond to different phases of the solar cycle.
We investigate the solar cycle dependence of $A$ and $\gamma$ in Figures \ref{fig:solarcycle_3} and \ref{fig:solarcycle_113}. 
The top and middle panels in the Figures \ref{fig:solarcycle_3} and \ref{fig:solarcycle_113} show the temporal variation of $\gamma$ and $A$. 
For the comparison, the yearly averaged sunspot number (SSN)\footnote{http://www.sidc.be/silso/datafiles} for different years are plotted in the bottom panel 
of Figures \ref{fig:solarcycle_3} and \ref{fig:solarcycle_113}.
Figure \ref{fig:solarcycle_3} corresponds to
$\alpha=3$ while Figure \ref{fig:solarcycle_113} refers to $\alpha = 11/3$. 
Upon comparing the top and middle panels with the bottom ones, it is evident that both $A$ and $\gamma$ are well correlated with the sunspot number. 
These trends hold irrespective of whether we use the proton gyroradius or proton inertial length prescription 
for the inner scale, and whether we use $\alpha = 11/3$ or $\alpha = 3$. 

The correlation between $A$ and the sunspot number is indicative of the fact that the overall magnitude of scattering is higher 
during solar maximum as compared to solar minimum. This is consistent with earlier results using interplanetary scattering observations 
\citep{Jan11, Man12, Jan15}. 

The correlation between $\gamma$ and the sunspot number indicates that the scattering strength falls off {\em faster} with heliocentric 
distance when solar activity increases. This might be because the large-scale solar magnetic field becomes more multipolar with increasing solar 
activity. For instance, this is reflected by the increasing complexity of the streamer belt with solar activity \citep{Wan00,Ric08}. 
Higher order multipolar fields are known to fall off more rapidly with heliocentric distance than a dipole, and this could be reflected in the spatial behaviour 
of the scattering strength, characterized by $\gamma$. Conversely, it has been reported earlier \citep{Tok00} 
that the scintillation index for IPS observations shows a rather shallow variation with heliocentric distance towards solar minimum. 
It should also be borne in mind that the Crab nebula passes from low latitudes to high(er) ones (upper panel of Figure \ref{fig:figure2}). Near solar minimum, 
this means that it progresses from sampling the slow solar wind to the fast solar wind, and this is an additional complicating factor. 
Near solar maximum, the solar wind is relatively more symmetric with latitude, and is predominantly slow \citep{Mcc2000, Asa1998}.
%

\begin{center}
\begin{sidewaystable}
\begin{tabular}{cccc|c|c|c|c|c|c|c}
\\\cline{1-11} \cline{1-11}

& Yearly 
& Observed
& \multicolumn{4}{c|}{Proton inertial length} 
& \multicolumn{4}{|c}{Proton gyroradius ($Ti=10^5$ K)}  \\\cline{4-11}

S.No
& averaged 
& Year 
& \multicolumn{2}{c|}{$\alpha=3$} 
& \multicolumn{2}{|c|}{$\alpha=11/3$}
& \multicolumn{2}{|c|}{$\alpha=3$} 
& \multicolumn{2}{|c}{$\alpha=11/3$}  \\\cline{4-11}

& sunspot number
& 
& $\gamma$
& A $(cm^{-6})$
& $\gamma$
& A $(cm^{-20/3})$
& $\gamma$
& A $(cm^{-6})$
& $\gamma$
& A $(cm^{-20/3})$  \\ \cline{1-11}


%

1&	261.7	&1958&	-4.2&	8.6E+6&	-4.9&	3.6E+3&	-3.7&	2.1E+5&	-4.7&	1.1E+3 \\
2&	200.7	&1956&	-4.9&	1.2E+8&	-5.6&	5.3E+4&	-4.4&	2.3E+6&	-5.4&	1.4E+4 \\
3&	159	&1960&	-2.7&	8.1E+4&	-3.3&	3.3E+1&	-2.3&	3.7E+3&	-3.2&	1.2E+1 \\
4&	94	&2013&	-3.4&	4.0E+5&	-3.7&	5.6E+1&	-2.7&	7.0E+3&	-3.9&	4.0E+1 \\
5&	80.8	&2011&	-2.8&	3.2E+4&	-3.2&	4.9E+0&	-2.1&	4.0E+2&	-3.0&	1.2E+0 \\
6&	76.4	&1961&	-2.8&	1.1E+5&	-3.5&	4.3E+1&	-2.6&	7.2E+3&	-3.4&	1.8E+1 \\
7&	54.2	&1955&	-4.9&	4.0E+7&	-5.6&	1.8E+4&	-4.2&	5.2E+5&	-5.3&	4.2E+3 \\
8&	53.4	&1962&	-2.2&	2.2E+3&	-2.8&	1.3E+0&	-2.5&	1.8E+3&	-2.9&	7.4E-1 \\
9&	45	&1952&  -2.1&	3.7E+4&	-2.7&	1.1E+1&	-1.7&	1.4E+3&	-2.6&	5.3E+0 \\
10&	20.1	&1953&	-2.7&	1.5E+5&	-3.4&	6.8E+1&	-2.3&	4.7E+3&	-3.2&	2.0E+1 \\
11&	6.6	&1954&	-2.9&	1.4E+5&	-3.6&	6.4E+1&	-2.8&	1.1E+4&	-3.6&	2.7E+1 \\
\cline{1-11}
\hline 
\end{tabular}
\parbox{17cm}{
\caption{$C_{N}^{2}$ as a function of heliocentric distance deduced from our observations.  
We fit the data for each year with a function of 
the form $C_N^2(R)=A~R^{-\gamma}$. This Table shows values for $A$ and $\gamma$.}
}
\label{tab:one}
\begin{minipage}{7cm}
\end{minipage}
\end{sidewaystable}
\end{center}

\subsection{Heliocentric and solar cycle dependence of $\epsilon_N \equiv \delta N_{k_{i}}/N$}
\begin{sidewaysfigure}
\includegraphics[width=23cm]{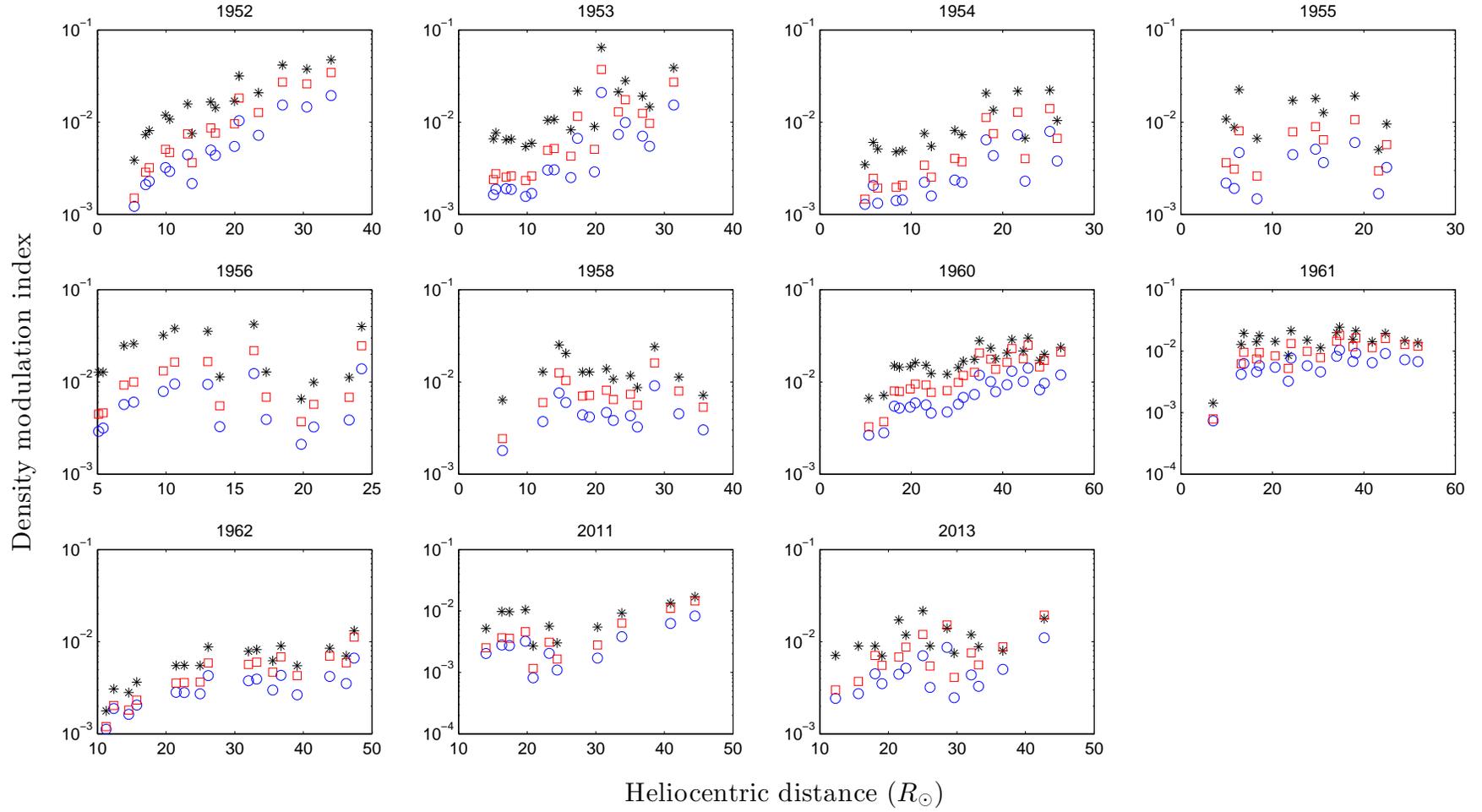}
\caption{The Figure shows the measured density fluctuation index ($\epsilon_N$) over a heliocentric 
distance in different years using different inner scale models. The `*' indicates the proton inertial length model and `circles' and `squares' represent 
the proton gyroradius model with proton temperature $10^5$ and $1.5 \times 10^6$ K respectively.}
\label{fig:dne_n}
\end{sidewaysfigure}

We next use our knowledge of $C_{N}^{2}$ to estimate the density modulation index $\epsilon_{N}$ using Eqs~(\ref{eq:deltn}) 
and (\ref{eq:df}). We use \citet{Leb98} prescription to evaluate the background solar wind density $N$.
The heliocentric distance dependence of $\epsilon_{N}$ is shown in Figure \ref{fig:dne_n} for different years. 
This quantity is computed using both the proton inertial length and proton gyroradius inner scale models. The broad conclusion 
that can be drawn from Figure \ref{fig:dne_n} is that $\epsilon_N$ ranges between 0.001 and 0.1, and its only weakly dependent
on heliocentric distance. The most we could discern was a linear dependence of $\epsilon_{N}$ with heliocentric distance with 
a slope of $1.45 \times 10^{-3}\,R_{\odot}^{-1}$ in 1952. During solar maximum years, however, the slope was close to zero. We note that 
\citet{Asa1998} have investigated the solar wind speed dependence of the density modulation index usig IPS observations;  this, in turn, can be related to solar cycle dependence.

\begin{figure*}
\centering
\includegraphics[width=12cm]{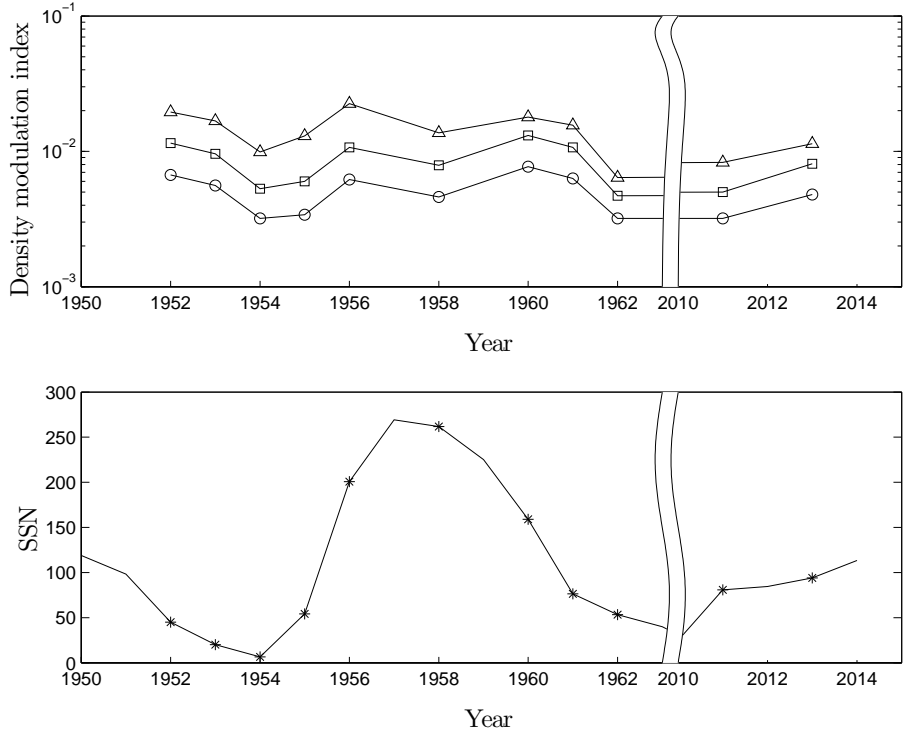}
\caption{The top panel of the Figure shows variation of the density fluctuations in different years is an
evidence of their dependence on solar cycle. The `triangles' indicate the proton inertial length model and the `circles' and `squares' 
indicate the proton gyroradius model with the proton temperature $10^5$ and $1.5 \times 10^6$ K respectively.
In the lower panel, `*' indicates the average sunspot number on corresponding years.}
\label{fig:epsilon}
\end{figure*}

Since the heliocentric distance dependence of $\epsilon_N$ is rather weak, it is meaningful to compute an average for this quantity for each year.
The average of $\epsilon_N$ between 10 and 45 $R_{\odot}$ is plotted as a function of time in the upper panel of Figure \ref{fig:epsilon}. 
Comparison with the lower panel, which shows the sunspot numbers, shows that $\epsilon_N$ broadly follows the solar cycle. 

However, we note that $\epsilon_N$ shows a prominent dip around 1958, which happens to be the year with the highest sunspot number of 
the data we have examined. Although the dip comprises only one data point, the following could be a tentative explanation for it:
\citet{Cel1987} notes that the modulation index ($\epsilon_N$) is positively correlated with the temperature of solar wind protons. At 1 AU, 
it is also observed that the proton temperature is positively correlated with solar wind speed \citep{Lop1986}.
Taken together, this implies that $\epsilon_N$ should be larger in the fast solar wind than in the slow solar wind. 
During the solar minimum, the Sun's large-scale magnetic field is predominantly dipolar. Consequently, higher latitudes 
are dominated by fast ($\approx 700 ~km/s$) solar wind emanating from coronal holes. Lower latitudes, on the other hand, 
are dominated by the slow solar wind ($\approx 400 ~km/s$) emanating from near the streamer belt. During solar maximum, however,
the large-scale solar magnetic fields is multi polar. Coronal holes are not as prevalent and slow solar wind is observed over all 
heliolatitudes \citep{Mcc2000, Asa1998}. Since 1958 was associated with a high sunspot number (the highest of the years we have 
considered), we expect slow solar wind (and low proton temperatures) at all heliolatitudes because the magnetic field is multipolar. 

Furthermore, \citet{Asa1998} suggest that the modulation index of the high speed solar wind (which is usually observed near solar minimum) 
shows significant evolution with heliocentric distance. Our results (Figure \ref{fig:dne_n}) show that the modulation index does not vary 
appreciably with heliocentric distance during the solar maximum years of 1956, 1958, 1960, 1961 and 2013, when the slow solar wind is expected 
to dominate. Our results are thus consistent with the converse of the conclusions reached by \citet{Asa1998}.

\section{Summary and Conclusions}

Density fluctuations are an important and relatively ill-understood facet of the phenomenon of solar 
wind turbulence. Most studies of solar wind turbulence in general, and density fluctuations in particular, 
concentrate on the spectral slope ($\alpha$, Eq~\ref{eq1}), and not so much on its spectral amplitude 
($C_{N}^{2}$, Eq~\ref{eq1}). Needless to say, the amplitude of the density turbulence spectrum is key 
to several important problems such as extended solar wind heating and angular broadening of radio sources. 
Our knowledge of $C_{N}^{2}$ (and its heliocentric dependence in particular) is currently limited to the 
investigations of \citet{Spa95} and \citet{Spa96} (quoted in Eq~\ref{eq2}), who used VLBI observations made 2 years past the maximum of cycle 19 
to probe scale sizes $\geq 200$ km. Their formulation used the Kolmogorov scaling ($\alpha = 11/3$) and did not consider an inner/dissipation scale. 
The density modulation index $\epsilon_{N}$ (Eq~\ref{eq:df}) is somewhat better studied. However, most of these studies have rather sparse coverage, 
and the only comprehensive study of this quantity that we are aware of \citet{Bis14b} is only for heliocentric distances $> 40 R_{\odot}$.

We use results from the standard technique of Crab nebula occultation to obtain a comprehensive palette of results 
concerning the heliocentric dependence of the density turbulence spectral amplitude ($C_{N}^{2}$) and the density
modulation index ($\epsilon_{N}$) for $10 < R < 45 R_{\odot}$. This is a distance range that is typically not covered 
by either IPS or interferometric techniques. We include the effects of the inner scale using currently prevalent models 
for it. Since the spatial scales used are small enough to possibly be comparable to the inner/dissipation scale, we use the general structure 
function (GSF) to model the observed visibilities rather than asymptotic approximations. Since there is evidence for 
flattening of the spectrum near the inner scale, we quote results for $\alpha = 11/3$ as well as a flatter value of 3. 
We parametrize the heliocentric dependence of the density turbulence amplitude as $C_{N}^{2}(R) = A\,R^{-\gamma}$; the values of $A$ and $\gamma$ from our observations are shown 
in Table~\ref{tab:one}. This gives an idea of the range of possibilities for the behavior 
of $C_{N}^{2}$ using currently prevalent ideas. To the best of our knowledge, this is the most extensive characterization of the density turbulence spectral amplitude to date. 
For example, for the proton 
inertial length prescription for the inner scale and $\alpha=3$, `A' ranges from $2.2 \times 10^3$ to $1.2 \times 10^8~ cm^{-6}$ and $\gamma$ ranges from $-2.1$ to $-4.9$.
With the same inner scale prescription, with $\alpha=11/3$, `A' varies between $1.3$ and $5.3 \times 10^4 ~m^{-20/3}$ and $\gamma$ ranges from -2.7 to -5.6.
With the proton gyroradius inner scale model and $\alpha=3$, `A' ranges from $4 \times 10^2 ~-~2.3 \times 10^6~ cm^{-6}$ and $\gamma$ ranges from -1.7 to -4.4. With the proton gyroradius inner scale model and $\alpha=11/3$, `A' ranges from $0.74$ to $1.4 \times 10^4~ cm^{-20/3}$ and $\gamma$ 
varies from -2.6 to -5.4. In the only instance where our results can be compared with the existing results of \citet{Spa95} and \citet{Spa96}, our values for $\gamma$ agree well with theirs. Given the widely different observational and theoretical interpretation techniques we use, and the fact that the observations we are using for comparison are from a different solar cycle, this is remarkable.

Since we have used data from varying stages of the solar cycle, we investigate the solar cycle dependence of $A$ and 
$\gamma$; the results for which are summarized in Figures~\ref{fig:solarcycle_3} and \ref{fig:solarcycle_113}. 
The behavior of $A$ confirms the well known fact that the overall strength of scattering increases with increasing
solar activity and vice-versa. Our results for $\gamma$ imply that the scattering amplitude decreases more rapidly 
with heliocentric distance with increasing solar activity. This is intriguing, and could reflect the increasingly multipolar nature of 
the large-scale coronal magnetic field near solar maximum, since higher order multipoles decay more rapidly with distance. Taken together, our results could have interesting implications for the connection 
between density fluctuations and the large scale solar magnetic field. The possible connection between declining (large-scale) polar fields and the 
density turbulence levels probed by the IPS technique has been pointed out earlier \citep{Jan10, Jan11, Jan15, Bis14a}. Our results are an 
interesting complementary take on this problem, using a different technique and for heliocentric distances that are much closer to the Sun.

We also use our knowledge of $C_{N}^{2}$ to obtain the density modulation index as defined in Eqs~(\ref{eq:deltn}) and (\ref{eq:df}). 
In agreement with the results of \citet{Bis14b} for larger heliocentric distances, we find that $\epsilon_{N}$ depends only weakly on
heliocentric distance. While \citet{Bis14b} found that $\epsilon_{N}$ shows a monotonic decline of around 8 \% over solar cycle 23, we
find that $\epsilon_{N}$ closely tracks the solar cycle, with a peak-to-peak variation (from 1956 to 1962) of around 72 \%. Our results 
on the density modulation index can be used to investigate some important questions regarding the solar wind: it can be used to calculate 
the extended solar wind heating rate, and it provides yet another way of investigating the relation between density turbulence, the large 
scale magnetic field and turbulent magnetic field fluctuations.

\acknowledgments

\begin{itemize}
 \item The Crab nebula occultation data for 2011 and 2013 is available on request from the Gauribidanur observatory, 
 operated by Indian Institute of Astrophysics, Bangalore, India. Email-id: ramesh@iiap.res.in.
 
  \item PS acknowledges support from the ISRO-RESPOND program. 
\item  The authors would like to thank the anonymous referees for their valuable and constructive suggestions.
\end{itemize}


%
%
%
%
%
%
%
%
%

%
%

\end{document}